# AN INTRODUCTION TO AGENT BASED MODELLING AND SIMULATION OF SOCIAL PROCESSES

**Armano Srbljinović[1,*] and Ognjen Škunca[2]**

[1]Croatian Ministry of Defence's Institute for R&D of Defence Systems
 Zagreb, Croatia

[2]Oikon Ltd.
 Zagreb, Croatia

## SUMMARY

The paper provides an introduction to agent-based modelling and simulation of social processes. Reader is introduced to the worldview underlying agent-based models, some basic terminology, basic properties of agent-based models, as well as to what one can and what cannot expect from such models, particularly when they are applied to social-scientific investigation. Special attention is given to the issues of validation.

*Corresponding author, $\eta$: armano.srbljinovic@kc.htnet.hr; +385 1 4603 840;
Ministry of Defence – Institute for R&D of Defence Systems, P.O. Box 252, HR - 10002 Zagreb, Croatia

## 1. INTRODUCTION

According to the modern scientific model of the world, the world that we live in is envisioned as a stratified structure with many levels [1]. Entities on any of the levels can be said that form, produce or make part of the entities constituting the next higher level[1]. Elementary particles form atoms. Atoms form molecules, simple and the more complex ones, and molecules form all objects in the universe. Among others, complex organic molecules form living cells. Cells form tissues and organisms. Living organisms form ecosystems. Neurones, a special kind of cells, form human brains capable of being self-conscious. Human beings form groups, societies and civilisations.

## 2. COMPLEXITY

Through identifying the key entities and their properties and modes of interaction, on each of these levels a framework of a particular scientific discipline has been established explaining the phenomena on the corresponding level. However, entities on each of these levels are not simple clusters or heaps of the lower level entities, not simple sums of their parts, but the more complex wholes, the interrelated and interactive structures possessing certain new characteristics and regularities[2]. Recognising that the traditional scientific disciplines are the least successful precisely in describing and explaining ways in which relatively simple parts organise or self-organise into more complex and sophisticated wholes, the relatively new discipline of *science of complexity*, or *complexity theory*, or shortly *complexity*, aims to tackle these traditionally insufficiently explored and understood issues [2]. Most generally complexity can be said to study *complex adaptive systems* (CAS) – dynamic systems consisted of many simple, typically nonlinearly interacting parts possessing capabilities of adaptation to their constantly changing environment. The main task for the science of complexity becomes to explain how relatively stable, aggregated, macroscopic patterns are induced by local interactions of multitudes of lower level entities.

## 3. AGENT-BASED SIMULATIONS

As these non-linear, adaptive interactions are mostly too complex to be captured by analytical expressions, computer simulations are most often used. The basic idea of such simulations is to specify the rules of behaviour of individual entities, as well as the rules of their interaction, to simulate a multitude of the individual entities using a computer model, and to explore the consequences of the specified individual-level rules on the level of population as a whole, using results of simulation runs. As the simulated entities are usually called *agents*, the simulations of their behaviour and interactions are known as *agent-based simulations*[3]. The properties of individual agents describing their behaviour and interactions are known as *elementary properties*, and the properties emerging on the higher, collective level are known as *emergent properties*.

## 4. BASIC PROPERTIES OF AGENT-BASED MODELS

What makes agent-based models particularly appealing and interesting is that consequences on the collective level are often neither obvious, nor expectable, even in many cases when the assumptions on individual agent properties are very simple. Namely, the capability of generating complex and intriguing emergent properties arises not so much from the in-built rules of individual agent behaviour, as from the complexity of the network of interactions among the agents. Precisely this multitude of agents, as well as the multitude and complexity of their interactions, are the main reasons why in most cases formal mathematical deduction of results of an agent-based model is not possible.

This is also the reason why the issues of complexity remained relatively under-explored until recently. Namely, as scientists regularly decide to pay attention to "problems defined by the conceptual and instrumental techniques already at hand" [3] (cited in [4]), "some facts […] are pushed to the periphery of scientific investigation, either because they are thought not to be relevant, or because their study would demand unavailable techniques" [4]. Accordingly, only after recent advances in the development of computational technology have enabled massive simulation experiments, the issues of emergent complexity came closer to the focus of scientific research.

Besides the above mentioned modelling of bottom-up effects, i. e. the effects originating at the individual level and influencing the collective one, more complex agent-based models are also capable of modelling top-down effects, arising at the collective level and influencing the level of individual agents.

## 5. AGENT-BASED MODELS IN SOCIAL SCIENCES

Although most well-established within the framework of natural sciences, the application of agent-based simulations within the field of social sciences since 1990-ies is also significantly growing.

It must be emphasised that the primary purpose of agent-based modelling and simulations in social sciences is not prediction. Namely, social processes are usually so complex that their maximally faithful replicas would hardly ever be possible. The consequence is that agent-based models mostly do not possess the level of "accuracy" needed for a model to be used for predictive purpose.

### 5.1. ADVANTAGES

The main purpose of agent-based models is to help in developing new and formalising already existing theories. Namely, with regard to the process of formalisation, which includes precise formulation of a theory, as well as securing its coherence and completeness, computer simulations in social sciences can be said to have the role similar to mathematics in natural sciences. As aspects that make computer simulations more appropriate for formalising social science theories than most of mathematical models (or than most of the "more elegant", closed-form mathematical expressions, at least), it is possible to identify the following ones [5]:

- programming languages are more expressive and less abstract than most mathematical techniques;
- computer programs deal more easily with parallel processes and processes without a well-defined order of actions than systems of mathematical equations;
- programs designed in accordance with the principles of software engineering are modular, which facilitates their modification; mathematical systems often lack this kind of modularity;
- it is easier to build simulation systems which include heterogeneous agents – for example, to simulate people with different perspectives on their social worlds, different stocks of knowledge, different capabilities and so on – while this is usually relatively difficult using mathematics.

Some additional advantages, more specific for agent-based models are [6]:

- possibility of modelling more "fluid" or "turbulent" social conditions when modelled agents and their identities are not fixed or given, but susceptible to changes that may include "birth" or "death" of individual agents, as well as adaptation of their behaviour;
- possibility of modelling boundedly rational agents, making decisions and acting in conditions of incomplete knowledge and information;
- possibility of modelling processes out of equilibrium.

The previously mentioned possibility of modelling top-down influences with agent-based models is also important when simulating social processes because human individuals are capable of observing patterns and emergent structures on the collective level, so that the existence of such patterns and structures often has a feedback effect on the behaviour of individuals. [5].

As already stated, agent-based models and simulations serve explanatory more than predictive purpose. They provide us with means of performing computer simulation-enhanced thought experiments aimed at improving our intuition about the modeled phenomena. This feature is particularly important in social sciences where possibilities of experimenting in real-world situations are rather limited. The results of thought experiments are to be contrasted with theory that was used when designing the experiment. "When a thought experiment generates dissonance (i.e., the consequences of the thought experiment are not easily accommodated by our current understanding of the phenomena involved) we must question both the integrity of our current theories, and the validity of the intuitions which guided our thoughts during the thought experiment. […] It may indeed be possible to make a stronger case with simulations than with a 'naked' thought experiment since a simulation can also provide insights that could not be arrived at by thinking alone." [4].

## 5.2.  WEAKER POINTS AND LIMITATIONS

However, this improved powerfulness and versatility of simulation-enhanced thought experiments comes at the price of "explanatory opacity," meaning that "[…] the behaviour of a simulation is not understandable by simple inspection; on the contrary effort towards explaining the results of a simulation must be expended, since there is no guarantee that what goes on in it is going to be obvious. […] Computer simulations must be observed and systematically explored before they are understood, and this understanding can be fed back into existing theoretical frameworks." [4].

Of course, this is easier said than done. Because of typically huge number of model parameters and a massive amount of model-generated data for each parameter configuration, the results are "fragile," meaning that it is often not easy to find out whether model results are a mere artefact of specific parameter configurations or the really meaningful results [6]. The theory underlying model's design may sometimes provide guidance as to which ranges of parameters are most critical to test, so that the total parameter space may be reduced to only a portion needing inspection.

"Step-by-step" method of design provides another way of reducing "explanatory opacity." "By restarting the analysis from scratch from time to time, and adding theoretical features incrementally, this design strategy makes it easier to manage overwhelming complexity" [6].

Because of the complexity of most agent-based models and, particularly, of their computer implementations, the models' communication is also often impaired, as well as reproducibility of results. To avoid some of these difficulties, it is recommended to use some of the already existing, more or less standardised software packages, designed with the specific purpose of facilitating design and development of agent-based models and simulations. On the other hand, clearly, packages are inevitably limited in what they can offer [5].

## 5.3. VALIDATION

By validation is here meant assuring external or operational validity which refers to the adequacy and accuracy of the model in matching real world data, where the real world data refer to information gathered through experimental, field, archival, or survey analyses of actual human, animal, physical systems, groups, or organisations [7]. The highly abstract nature of agent-based modelling makes validation of such models difficult. How can we "infer emergence as a causal mechanism in the real world, once we have identified it in the CAS? […] It is possible that the causal mechanism hinted at in the CAS is swamped by the additional 'turbulence' in the real world, and some entirely different sets of interactions of direct effects drive the formation of the feature of interest." [8] (cited in [6]).

Some researchers, particularly in the area of so-called artificial societies, avoid this issue by stating as their goal finding theories which apply not just to human societies, but to societies of interacting agents generally [5]. They view their activity as an "attempt to grow certain social structures in the computer – or in silico – the aim being to discover fundamental local or micro mechanisms that are sufficient to generate the macroscopic social structures and collective behaviours of interest." [9].

Still other researchers point out that positivist tests of specific predictions are not appropriate to find out whether the postulated processes operate in the real world because "[…] the CAS approach produces 'pattern predictions' or 'robust processes' rather than point-like predictions of single events" [6]. The term "robust process" refers to "a sequence of events that has unfolded in similar (but neither identical nor fully predictable) fashion in a variety of different historical contexts" [10] (cited in [6]). In order to check the external validity of robust processes, "dynamic understanding" is needed – "understanding that is holistic, historical, and qualitative, eschewing deductive systems and causal mechanisms and laws" [11] (cited in [6]). While some researchers, particularly those coming from the tradition of natural sciences, may find this position too close to mysticism for their scientific taste, one cannot deny that validation of agent-based models of social processes inevitably assumes a degree of arbitrariness and subjective judgement[5].

## 5.4. VALIDATION AS "GROUNDING"

Carley also argues that for "intellective models", aimed "to show proof of concept or to illustrate the relative impact of basic explanatory mechanisms, […] validation is somewhat less critical," the most important being "to keep a balance between keeping a model simple and attaining veridicality" [7]. As agent-based models in social sciences, which are used for purposes of thought experiments, may be said to mostly belong to the class of intellective models, techniques used for intellective models seem also appropriate for most agent-based models. The following account of such techniques closely follows Carley [7].

Stating the essential motto of intellective models as "keep it simple," Carley observes the importance of establishing that the simplifications made in designing the model do not seriously detract from its credibility and the likelihood of providing important insights. The process of "grounding" usually does this.

There are at least three aspects of grounding. The first one is "story telling" consisting of setting forth a claim for why the proposed model is reasonable, and then enhancing this claim by not overclaiming the applicability of the model and by discussing the model's limitations and scope conditions. This enhancement can be done in several ways: by demonstrating that other researchers have made similar or identical assumptions in their models; by explaining how the proposed model extends, is a special case of, is a generalisation of, or competes with one or more other models; and by demonstrating that the proposed model captures the key elements of a specific group, organisational, or social process, or the core ideas in a verbal theory.

We may add to this that it is indeed advisable to justify each part of the model and each modelling construct by showing how it derives from its corresponding theoretical roots. However, even when a well-formulated theory already exists, translating this theory from a human to a programming language may pose difficulties. As known very well, a computer model requires very precise values of model parameters and rigorously specified rules of variable manipulations. This level of precision most of theories do not meet, which, by the way, does not need to prevent them from being very successful in explaining modelled phenomena. Therefore "stipulative patches" are often needed when translating theories into computational models [12]. These patches present clearly articulated assumptions, necessary to implement the model on a computer, but derived more on a common sense, ad hoc basis, rather than the basis of a well founded theory. Clearly, such patches should be used only in the absence of the more reliable theoretical underpinnings.

Returning to the discussion of aspects of grounding, as the second aspect Carley lists initialisation – setting the various parameters and procedures so that they match real data. Finally, the third aspect of grounding is "performance evaluation" – the process of determining whether the model generates the stylised results or behaviour expected of the underlying processes. Clearly, such "non-surprising results" are generally not, and should not be, the only results that can be generated from the model, but establishing these results first is also a form of validation.

## 6. INSTEAD OF A CONCLUSION

After listing some of the advantages, as well as the difficulties associated with agent-based modelling, we would like to simply conclude by observing that the area of application of agent-based models and simulation in social sciences is rapidly expanding, spreading from psychology, anthropology and sociology, over economy and organisational theory, to political science [13].

Within the applications in international relations only, several lines of research may be identified: research and development of artificial geopolitical systems [6 (pp. 72-135), 14 – 16, 17 (pp. 121-144)] development of models of conflict based in game theory and extending them using agent-based models [17 (pp. 44-68), 18] use of agent-based models in exploring identity issues [6 (pp. 184-212), 12, 19].

## 7. REMARKS

[1]The list of examples that follows is by no means definitive or exhaustive, but serves the intended illustrative purpose.

[2]The multilevel flexibly co-ordinated structures that despite their complexity act as wholes are sometimes called holarchies [1].

[3]We use terms "agent-based simulations" and "agent-based models" intermittently although "model" usually denotes any representation of a part of reality, while "simulation" is more often used for a model representing how a part of reality changes in time, or for an execution of such a model. Agent-based models are often also called "multi-agent models."

[4]Among the most popular, freely available software packages providing support for design and development of agent-based computer simulations are *SWARM* (www.swarm.org) and *RePast* (http://repast.sourceforge.net/).

[5]Instead of term "validation", the term "empirical evaluation" is often used to emphasise the relaxed character of validating social-scientific agent-based models, as opposed to the more demanding validation of engineering models of technical systems.

# UVOD U MODELIRANJE POMOĆU AGENATA I SIMULACIJU DRUŠTVENIH PROCESA

Armano Srbljinović[1] i Ognjen Škunca[2]

[1]MORH – Institut za istraživanje i razvoj obrambenih sustava,
Zagreb, Hrvatska

[2]Oikon d.o.o.,
Zagreb, Hrvatska

## SAŽETAK

Članak donosi uvod u modeliranje i simulacije društvenih procesa temeljene na agentima. Čitatelja se uvodi u svjetonazor u pozadini modela temeljenih na agentima, osnovnu terminologiju, osnovna svojstva modela temeljenih na agentima, te u ono što se može i što se ne može očekivati od takvih modela, posebno kad ih se primjenjuje u društvenoznanstvenim istraživanjima. Posebna pažnja pridana je pitanjima validacije.

## KLJUČNE RIJEČI

modeli temeljeni na agentima, simulacije društvenih procesa, teorija kompleksnosti, kompleksni adaptivni sustavi